\documentclass[12pt]{article}
\title{Quantized Fields in a Nonlinear Dielectric Medium: A Microscopic 
Approach}\author{Mark Hillery\\Department of Physics and Astronomy
\\Hunter College of the City University of New York\\695 Park Ave.
\\New York, NY 10021\and Leonard Mlodinow\\Disney Interactive\\Mail Code
8161\\500 S.\ Buena Vista St.\\Burbank, CA 91521}

\begin{document}
\maketitle
\begin{abstract}
Theories which have been used to describe the quantized electromagnetic 
field interacting with
a nonlinear dielectric medium are either phenomenological or derived
by quantizing the macroscopic Maxwell equations.  Here we take a different 
approach and derive a Hamiltonian describing interacting fields from from one
which contains both field and matter degrees of freedom.  The medium is 
modelled as a collection of two-level atoms, and these interact with the 
electromagnetic field.  The atoms are grouped into effective spins and the
the Holstein-Primakoff representation of the spin operators is used to expand 
them in inverse powers of the total spin.  
When the lowest order term of the interaction is combined with the
free atomic and field Hamiltonians, a Hamiltonian describing a theory of 
noninteracting polaritons results.  When higher order terms are expressed in
terms of polariton operators standard nonlinear optical interactions emerge.
These are then compared to the results of phenomenological and macroscopic 
theories.  The theory is also used to derive an effective Hamiltonian describing 
counterpropagating modes in a nonlinear medium.
\end{abstract}
\pagebreak
\section{Introduction}
The study of the propagation of quantized electromagnetic fields in nonlinear 
dielectric media
is an area in which there is presently considerable theoretical and experimental
activity [1,2].  This interest is being driven by both fundamental and practical
considerations.  On a fundamental level nonlinear optics provides an example of 
an experimentally accessible nonlinear quantum 
field theory, and the light which is produced
can exhibit distinctly quantum effects such as squeezing and quantum phase
 diffusion [3-7].  The practical 
interest comes from the proposed use of optical solitons in fibers for long-
distance communication [8]. This has made necessary a better understanding 
of the role of quantum noise in such systems.

The first step which must be taken in these problems is the quantization 
of the electromagnetic
field in the presence of a nonlinear dielectric.  This can be done in two ways
which we shall call the macroscopic and microscopic approaches.  In the first,
the macroscopic approach, the medium is completely described by its linear and
nonlinear susceptibilities.  No matter degrees of freedom appear explicitly in
this treatment.  One then finds a Lagrangian which produces the macroscopic
Maxwell equations for the field in a nonlinear medium.  From it one finds the 
canonical momenta and the Hamiltonian. Quantization is accomplished by
imposing the standard equal-time commutation relations.  We explored 
this approach was in a previous paper [9].  In the second approach, 
the microscopic, a model for the medium is constructed, 
and both the field and matter degrees of freedom appear in the theory.  
Both are quantized.  This is what we shall explore here.  
As we shall see, the result is a theory of mixed matter-field modes which are 
coupled by a nonlinear interaction.

A number of problems arise in the quantized macroscopic theory 
which lead us to examine a microscopic one.  The first is the difficulty
in including 
dispersion.  Dispersion is a result of the fact that the medium is nonlocal
in time, i. e. the polarization at a given time depends not only on the 
electric field at that time but also on the field at earlier times.  It is not
clear whether nonlocal theories can be encompassed within the 
canonical formalism.  (Drummond has shown how it can be done if one is
considering a narrow frequency interval [10].  He then showed that additional
intervals could be considered at the expense of introducing additional fields.)
Another issue is operator ordering.  Because the vacuum state of the theory now
includes the effects of the medium, it is no longer clear that normal ordering
is the appropriate form to use.  Finally, it is expected that the inclusion
of losses in the theory will be most easily accomplished at the 
microscopic level [11].

The macroscopic and microscopic approaches are complementary; 
each has its own range of application and illuminates the other.  The 
microscopic theory we present here, being more fundamental, 
is more appropriate 
for addressing basic problems such as dispersion and operator ordering.  It is,
however, at the moment restricted to rather simple models of the medium.  Real
media are more easily described by the macroscopic theory in that the medium is
characterized by only a few parameters.  It has been used, for example, by
Drummond and Carter to study the propagation of quantized fields in silica
fibers [4].

While Drummond and Carter have developed the macroscopic theory into a useful
tool, much of the work done so far on the nonlinear optics of quantized fields 
has used a phenomenological approach.  
One assumes that the field in a $\chi^{(3)}$ 
medium is described by the nonlinear Schroedinger equation and the theory is
quantized by imposing equal-space, rather than equal-time, commutation 
relations [5].
A study of the use of equal-space commutations has been made by Deutsch [12].  
He 
concluded that for linear theories they lead to the same results as the more
standard equal-time relations, but for nonlinear theories it is expected that
for some quantities this will no longer be the case.  These issues are 
in need of further clarification.

We shall consider a medium consisting of two-level atoms.  This is a model
which we treated before, but in that case the field was restricted to a single
mode [13].  This did not allow the inclusion of propagation.  
We now extend that work
to include an arbitrary number of modes.  This extension is not straightforward, 
because in the single-mode case all of the two-level atoms can be treated as
one large spin.  This is no longer possible when more than one mode is present.  
This necessitates modification of the methods used in our first paper.  To
accomplish this, we divide the medium into little boxes whose dimensions are
smaller than the shortest wavelength which appears in the theory.  The atoms
in each of these boxes can be treated as a large spin and the Holstein-
Primakoff representation of the resulting spin operators can then be used.
This allows us to make a $1/N$ expansion of the Hamiltonian where $N$ is the
number of atoms in a box [14].  A continuum version of the theory is obtained by
letting the box size go to zero.

What results is a Hamiltonian containing electromagnetic and matter fields.  The 
linear part of the Hamiltonian is essentially the same as that obtained 
by Hopfield
in his treatment of linear dielectric media [15].  It can be diagonalized by 
transforming from the matter and field modes to mixed matter-field modes, known
as polaritons.  When the nonlinear part of the 
Hamiltonian is expressed in terms
of polariton operators the result consists 
of interaction terms which are similar
to those which are familiar from nonlinear optics, i.\ e.\ terms describing an
intensity-dependent index of refraction and third harmonic generation. 
The effects
of dispersion are automatically included because 
the polaritons have a dispersion
relation which differs from that of photons.  The operator ordering in the 
Hamiltonian is also dictated by the theory.  The end result is a microscopic
theory to which the macroscopic and phenonenological theories can be compared.  
The microscopic theory is derived in Sections II-IV and the comparisons with
other theories are made in Section V.

\section{Hamiltonian for a Medium of Two-Level Atoms}
Let us consider a large number, $N$, of atoms 
distributed uniformly throughout a
volume $V$.  These atoms interact with the radiation field whose quantization 
volume is also taken to be $V$.  
For simplicity we shall assume that each atom has
a single optically active electron.  The Hamiltonian describing this system is
\begin{equation}
H = \sum_{j=1}^{N} \left[\frac{1}{2m}(\vec{p}_{j}+e\vec{A}(\vec{x}_{j}))^{2}
+V(\vec{x}_{j}-\vec{R}_{j})\right] +\frac{1}{2} \int d^{3}r [\dot{\vec{A}}^{2}
+ (\nabla \times \vec{A})^{2}].
\end{equation}
Here $\vec{x}_{j}$ and $\vec{p}_{j}$ are 
the position and momentum of the electron
on the jth atom which is located at $\vec{R}_{j}$.  The potential energy of the 
electron in the jth atom is $V(\vec{x}_{j}-\vec{R}_{j})$.  The charge on the 
electron is $-e$ where $e>0$.  Finally, we are assuming that we are in the 
Coulomb gauge so that $\nabla \cdot \vec{A} = 0$.

We now make two approximations.  First we wish to specialize to the case where 
only two levels for each atom are included;
an upper level $|a\rangle$ with energy $E_{0}/2$ and a lower level $|b\rangle$
with energy $-E_{0}/2$.  The projection operator onto the two-state space is
\begin{equation}
P= \prod_{j=1}^{N} (|a_{j}\rangle \langle a_{j}| 
+ |b_{j}\rangle \langle b_{j}|),
\end{equation}
and the effective Hamiltonian for the two-level atoms is obtained by applying 
it to both sides of Eq.(1).  Our Hamiltonian now becomes
\begin{eqnarray}
H & = & P\left\{\sum_{j=1}^{N}\left[\frac{1}{2}E_{0}\sigma_{j}^{(3)}+
\frac{e^{2}}{2m}
\vec{A}(\vec{x}_{j})^{2}\right] + \frac{1}{2} \int d^{3}r [\dot{\vec{A}}^{2}
+ (\nabla \times \vec{A})^{2}]\right\}P \nonumber \\ 
 & & \mbox{}+ \frac{e}{m} P\left[\sum_{j=1}^{N}
\vec{p}_{j}\cdot \vec{A}(\vec{x}_{j})\right] P.
\end{eqnarray}
Because we are dealing with optical wavelengths we are also in a position to
make the dipole approximation and replace $\vec{x}_{j}$, 
the electron coordinate,
by $\vec{R}_{j}$, the position vector of the atom, in the argument of $\vec{A}$.
This yields
\begin{eqnarray}
H & = & P\left\{\sum_{j=1}^{N}\left[\frac{1}{2}E_{0}\sigma_{j}^{(3)}+
\frac{e^{2}}{2m}
\vec{A}(\vec{R}_{j})^{2}\right] + \frac{1}{2} \int d^{3}r [\dot{\vec{A}}^{2}
+ (\nabla \times \vec{A})^{2}]\right\}P \nonumber \\
 & & \mbox{} + \frac{e}{m} P\left[\sum_{j=1}^{N}
\vec{p}_{j}\cdot \vec{A}(\vec{R}_{j})\right] P.
\end{eqnarray}

Let us now examine the last term of the above equation 
which represents the dipole
interaction between the atoms and the field.  As a first step we expand the
vector potential in plane wave modes
\begin{equation}
\vec{A}(\vec{R}) = \sum_{\vec{k},\lambda}
\sqrt{\frac{1}{2\omega_{k}V}}(a_{\vec{k},
\lambda}e^{i\vec{k}\cdot\vec{R}}+a_{\vec{k},\lambda}^{\dagger} e^{-i\vec{k}
\cdot\vec{R}})\hat{\epsilon}_{\lambda}(\vec{k}),
\end{equation}
where $\omega_{k}=|\vec{k}|$ and the polarization vectors, 
$\hat{\epsilon}_{\lambda}
(\vec{k})$, where $\lambda = 1,2$, satisfy $\vec{k}\cdot\hat{\epsilon}_{\lambda}
(\vec{k}) =0$ and are real.  We also adopt the convention
\begin{equation}
\hat{\epsilon}_{\lambda}(-\vec{k}) = (-1)^{\lambda}
\hat{\epsilon}_{\lambda}(\vec{k}).
\end{equation}
If we now define 
\begin{equation}
i\mu_{\lambda}(\vec{k})=\frac{e}{m} 
\langle a|\vec{p} |b\rangle\cdot \hat{\epsilon}
_{\lambda}(\vec{k})= ieE_{0}\langle a|\vec{x} |b\rangle\cdot \hat{\epsilon}
_{\lambda}(\vec{k}),
\end{equation}
then the interaction term becomes
\begin{eqnarray}
\frac{e}{m} \sum_{j=1}^{N}P[\vec{A}(\vec{R}_{j})\cdot\vec{p}_{j}]P & = & 
P\sum_{j=1}^{N}
\sum_{\vec{k},\lambda} 
\sqrt{\frac{1}{2\omega_{k}V}}(a_{\vec{k}\lambda}e^{i\vec{k}\cdot
\vec{R}_{j}}+a_{\vec{k}\lambda}^{\dagger}e^{-i\vec{k}\cdot\vec{R}_{j}})
\nonumber \\
 & & i(\mu_{\lambda}(\vec{k})\sigma^{(+)}_{j}-\mu_{\lambda}^{\ast}(\vec{k})
\sigma^{(-)}_{j})P
\end{eqnarray}

At this point we shall make a number of simplifications. 
First, we shall not write
the projection operators explicitly and shall assume that all matter operators 
act only on the two levels, $|a\rangle$ and $|b\rangle$, of each atom.  
We shall also assume that the phases of the atomic 
wave functions have been chosen so that 
$\mu_{\lambda}(\vec{k})$ is real.  Finally, 
we shall look at the case when only one polarization 
of the field is significantly populated so that 
the polarization index,$\lambda$,
will be dropped.  The remaining polarization will be assumed to 
obey Eq. (6) with $\lambda = 2$.

We now wish to rephrase the theory in terms of effective spins.  To this end
we divide the total volume, $V$, into blocks of volume $\Delta V$.  
The center of the 
$l$th block is located at the position $\vec{r}_{l}$.  
We suppose that there are
$n_{0}$ atoms in each block, where $n_{0}>>1$, and that the dimensions of each
block are much smaller than an optical wavelength.  This means that if we are
looking at optical phenomena all of the atoms in the $l$th block can be treated
as if they are at $\vec{r}_{l}$.  The atoms can then be treated as a spin $s=
n_{0}/2$ object located at $\vec{r}_{l}$.  Expressing the Hamiltonian in terms
of the block variables we have
\begin{equation}
H=\sum_{l=1}^{N_{b}}E_{0}S^{(3)}_{l} +\sum_{\vec{k}}
\omega_{k}a^{\dagger}_{\vec{k}}
a_{\vec{k}}+\frac{e^{2}n_{0}}{2m}\sum_{l=1}^{N_{b}}\vec{A}(\vec{r}_{l})^{2}
+H_{int},
\end{equation}
where
\begin{equation}
H_{int}=\sum_{l=1}^{N_{b}}\sum_{\vec{k}}\sqrt{\frac{1}{2\omega_{k}V}}
(a_{\vec{k}}e^{i\vec{k}\cdot\vec{r}_{l}}
+a_{\vec{k}}^{\dagger}e^{-i\vec{k}\cdot\vec{r}_{l}})
i\mu(\vec{k})(S^{(+)}_{l}-S^{(-)}_{l}).
\end{equation}
We have designated the $z$ component of the spin at 
$\vec{r}_{l}$ by $S^{(3)}_{l}$ and
the raising and lowering operators by $S^{(+)}_{l}$ and $S^{(-)}_{l}$, 
respectively.
$N_{b}$, which is equal to $V/\Delta V$, is the total number of blocks.

We can further simplify the $\vec{A}^{2}$ term in the Hamiltonian.  
If we multiply
and divide the sum by $\Delta V$ we have
\begin{eqnarray}
\frac{e^{2}}{2m}\left(\frac{n_{0}}{\Delta V}\right) \Delta V\sum_{l=1}^{N_{b}}
\vec{A}(\vec{r}_{l})^{2} &\cong &\frac{e^{2}\rho}{2m}\int_{V}d^{3}r\vec{A}
(\vec{r})^{2} \nonumber \\
 & \cong & \frac{e^{2}\rho}{2m}\sum_{\vec{k}}\frac{1}{2\omega_{k}}[a_{\vec{k}}
a_{-\vec{k}}+a_{\vec{k}}a_{\vec{k}}^{\dagger} \nonumber \\
 & & \mbox{}+a_{\vec{k}}^{\dagger}a_{\vec{k}}
+a_{\vec{k}}^{\dagger}a_{-\vec{k}}^{\dagger}].
\end{eqnarray}
We have denoted by $\rho$ the density of atoms which is equal to either $N/V$
or $n_{0}/\Delta V$.

We now have our Hamiltonian in the desired form.  The medium is described by
spin variables and the field by creation and annihilation operators.  
The next step is to expand it so that
we can extract the linear interaction and the different orders of nonlinear
interaction.  We accomplish this with a semiclassical,
 or $1/s$, expansion.

\section{Expansion and Continuum Limit}
Before proceeding to the actual expansion we must 
restrict the $\vec{k}$
summations by imposing a high-frequency cutoff.  
This is necessary 
so that the wavelengths in the theory do not become 
so small that they violate the 
conditions under which the macroscopic theory is valid.  In particular, we have 
assumed that the mean spacing between atoms is much less than any wavelength in
the theory.  Therefore, we shall restrict $|\vec{k}|$ to be less than $k_{u}$ 
where $k_{u}$ corresponds to a wavelength shorter 
than those in the optical regime
but considerably larger than the interatomic separation.

We would expect that the ground state of 
the field-atom system would be the state 
in which all of the spins are down, corresponding to all of the atoms in their
ground states, and no photons present.  
This is actually the case in the semiclassical
approximation.  A derivation of this fact is presented in Appendix A.  Here
we shall assume this to be the case and shall expand our Hamiltonian about the 
no-photon, all-spin-down state.

In the case of the spin operators this expansion is implemented by means of the 
Holstein-Primakoff representation [16] in which the spin $s$ operators 
$S^{(3)}$, $S^{(+)}$,
and $S^{(-)}$ are represented in terms of boson creation and annihilation
operators, $\zeta^{\dagger}$ and $\zeta$ as
\begin{eqnarray}
S^{(-)}=(2s-\zeta^{\dagger}\zeta )^{1/2}\zeta 
&\hspace{1cm}& S^{(+)}=\zeta^{\dagger}
(2s-\zeta^{\dagger}\zeta )^{1/2} 
\nonumber \\ & S^{(3)}=-s+\zeta^{\dagger}\zeta , & 
\end{eqnarray}
where $[\zeta ,\zeta^{\dagger}]=1$.  
Our convention for the square roots is that
when the argument is positive so is the square root, and when the argument is
negative, the square root is $i$ times a positive number.  
Eqs. (12) then give us
the proper commutation relations for the spin operators, i.\ e.\ 
\begin{equation}
[S^{(3)},S^{(\pm)}]=\pm S^{(\pm)} \hspace{1cm} [S^{(+)},S^{(-)}]=2S^{(3)}.
\end{equation}

The excitation number for the boson operators, i.\ e.\ the eigenvalue of 
$\zeta^{\dagger}\zeta$,
corresponds to $s_{3}+s$ where $s_{3}$ is the eigenvalue of $S^{(3)}$.  
Therefore, the boson vacuum state is the spin 
state with the spin pointing down.  If we are 
only considering states whose excitation number is small then we can expand the
square roots
\begin{equation}
S^{(-)} \cong \sqrt{2s}\left( 1-\frac{1}{4s} \zeta^{\dagger}\zeta\right) \zeta 
\hspace{1cm}
S^{(+)}=\sqrt{2s}\zeta^{\dagger}\left( 1-
\frac{1}{4s} \zeta^{\dagger}\zeta\right).
\end{equation}
In our model of a nonlinear medium the fraction of atoms in a block which is 
excited will be small because we are off resonance.  This corresponds to small
excitation numbers.  
Therefore, the use of the expansions in Eq. (14) is justified.
With these approximations our Hamiltonian becomes
\begin{equation}
H=H_{0}+H_{int}^{(1)}+H_{int}^{(2)},
\end{equation}
where
\begin{eqnarray}
H_{0}& = & \sum_{l=1}^{N_{b}}E_{0}(-s+\zeta^{\dagger}_{l}\zeta_{l})+
\sum_{|\vec{k}|<k_{u}}
\omega_{k}a_{\vec{k}}^{\dagger}a_{\vec{k}} +\frac{e^{2}\rho}{2m}\int_{V}d^{3}r
\vec{A}(\vec{r})^{2} \\
H_{int}^{(1)} &= & i\sum_{l=1}^{N_{b}} 
\sum_{|\vec{k}|<k_{u}}\sqrt{\frac{s}{\omega_{k}V}}
\mu(\vec{k})(a_{\vec{k}}e^{i\vec{k}\cdot\vec{r}_{l}}+a_{\vec{k}}^{\dagger}
e^{-i\vec{k}\cdot\vec{r}_{l}})(\zeta^{\dagger}_{l}-\zeta_{l}) \\
H_{int}^{(2)} & = & -\frac{i}{4s}\sum_{l=1}^{N_{b}} \sum_{|\vec{k}|<k_{u}}
\sqrt{\frac{s}{\omega_{k}V}}
\mu(\vec{k})(a_{\vec{k}}e^{i\vec{k}\cdot\vec{r}_{l}}
+a_{\vec{k}}^{\dagger}e^{-i\vec{k}\cdot\vec{r}_{l}}) \nonumber \\
 & & ((\zeta^{\dagger}_{l})^{2}\zeta_{l}-\zeta_{l}^{\dagger}\zeta_{l}^{2}),
\end{eqnarray}
and $[\zeta_{l},\zeta_{l^{\prime}}^{\dagger}]=\delta_{ll^{\prime}}$.  
$H_{int}^{(1)}$ 
represents the linear part of the interaction between the atoms and the field
while $H_{int}^{(2)}$ represents the nonlinear interaction.

We now want to go to a continuum representation of the matter operators.  
Instead of
the operators $\zeta_{l}$ and $\zeta_{l}^{\dagger}$ 
we wish to employ operators,
$\zeta(\vec{r})$ and $\zeta^{\dagger}(\vec{r})$ 
which are functions of a continuous
position variable and whose commutation relations are
\begin{equation}
[\zeta(\vec{r}),\zeta^{\dagger}(\vec{r}\,^{\prime})]=
\delta^{3}(\vec{r}-\vec{r}\,^{\prime}).
\end{equation}
Note that the operators $\zeta_{l}/\sqrt{\Delta V}$ and 
$\zeta_{l^{\prime}}^{\dagger}/
\sqrt{\Delta V}$ have the commutation relations
\begin{equation}
\left[ \frac{1}{\sqrt{\Delta V}}\zeta_{l},\frac{1}{\sqrt{\Delta V}}
\zeta_{l^{\prime}}^{\dagger}
\right] = \frac{1}{\Delta V} \delta_{ll^{\prime}},
\end{equation}
which in the limit $\Delta V \rightarrow 0$ limit goes to Eq. (19).  
This gives us the identification
\begin{equation}
\frac{1}{\sqrt{\Delta V}}\zeta_{l} \rightarrow \zeta(\vec{r}).
\end{equation}

This relation can be used to find the continuum 
representation of the different terms in
the Hamiltonian.  For example, the $\zeta_{l}^{\dagger}\zeta_{l}$ term becomes
\begin{equation}
\sum_{l=1}^{N_{b}}\zeta_{l}^{\dagger}\zeta_{l}=
\sum_{l=1}^{N_{b}}\frac{1}{\sqrt{\Delta V}}
\zeta_{l}^{\dagger}\frac{1}{\sqrt{\Delta V}}\zeta_{l}\Delta V \rightarrow
\int_{V}d^{3}r\zeta^{\dagger}(\vec{r})\zeta(\vec{r}),
\end{equation}
which, after noting that $N=2sN_{b}$, gives us
\begin{eqnarray}
H_{0} & = & -\frac{1}{2}NE_{0}+E_{0}
\int_{V} d^{3}r\zeta^{\dagger}(\vec{r})\zeta(\vec{r})
+\sum_{|\vec{k}|<k_{u}}\omega_{k}a_{\vec{k}}^{\dagger}a_{\vec{k}} \nonumber \\
 & & \mbox{}+\frac{e^{2}\rho}{2m}
\sum_{|\vec{k}|<k_{u}}\frac{1}{2\omega_{k}}[a_{\vec{k}}a_{-\vec{k}}+
a_{\vec{k}}a_{\vec{k}}^{\dagger}+a_{\vec{k}}^{\dagger}a_{\vec{k}}+
a_{\vec{k}}^{\dagger}a_{-\vec{k}}^{\dagger}].
\end{eqnarray}
Similarly, for $H_{int}^{(1)}$ and $H_{int}^{(2)}$ we find 
\begin{eqnarray}
H_{int}^{(1)} & = & \frac{i}{\sqrt{V}}\sum_{|\vec{k}|<k_{u}}
\int_{V} d^{3}r(a_{\vec{k}}
e^{i\vec{k}\cdot\vec{r}}+a_{\vec{k}}^{\dagger}e^{-i\vec{k}\cdot\vec{r}})
g_{\vec{k}}
(\zeta^{\dagger}(\vec{r})-\zeta(\vec{r})) \\
H_{int}^{(2)} & = & -\frac{i}{2\rho\sqrt{V}}\sum_{|\vec{k}|<k_{u}}
\int_{V} d^{3}r(a_{\vec{k}}
e^{i\vec{k}\cdot\vec{r}}+a_{\vec{k}}^{\dagger}e^{-i\vec{k}\cdot\vec{r}})
g_{\vec{k}}
(\zeta^{\dagger}(\vec{r})^{2}\zeta(\vec{r}) \nonumber \\
 & & -\zeta^{\dagger}(\vec{r})\zeta(\vec{r})^{2}),
\end{eqnarray}
where $g_{\vec{k}}=\mu(\vec{k})\sqrt{\rho/2\omega_{k}}$.

\section{Diagonalization of the Quadratic Part}
So far we have derived a Hamiltonian which 
contains two interacting fields, the
electromagnetic field and the matter field.  These fields have both linear and
nonlinear interactions with each other.  
We have yet to see an interaction of the
type found in nonlinear optics, that of a 
field interacting nonlinearly with itself.
For example, it is not immediately clear how 
something that looks like the usual
description of self-phase modulation would be described by our Hamiltonian.

Something which does look more like the usual theory emerges if we diagonalize 
the quadratic part of the Hamiltonian, i.\ e.\ $H_{0}+H_{int}^{(1)}$.  
What results
is a description in terms of mixed matter-field modes, or polaritons [15].  If
$H_{int}^{(2)}$ is expressed in terms of polariton operators, it describes a
nonlinear interaction between polaritons.  Therefore, instead of a Hamiltonian
describing the interaction of a photon field with itself, 
which phenomenological effective
Hamiltonians in nonlinear optics do, we have a Hamiltonian which describes the
interaction of a polariton field with itself.

In order to diagonalize $H_{0}+H_{int}^{(1)}$ we introduce the operators
\begin{equation}
\zeta_{\vec{k}}=\frac{1}{\sqrt{V}}
\int_{V}d^{3}re^{-i\vec{k}\cdot\vec{r}}\zeta(\vec{r}),
\end{equation}
which implies that
\begin{equation}
\zeta(\vec{r})=\frac{1}{\sqrt{V}}
\sum_{\vec{k}}e^{i\vec{k}\cdot\vec{r}}\zeta_{\vec{k}}.
\end{equation}
These operators have the comutation relations
\begin{equation}
[\zeta_{\vec{k}},\zeta_{\vec{k}^{\prime}}^{\dagger}]=
\delta_{\vec{k},\vec{k}^{\prime}}.
\end{equation}
In terms of these operators $H_{0}+H_{int}^{(1)}$ become
\begin{eqnarray}
H_{0} & = & -\frac{1}{2}NE_{0}+E_{0}\sum_{|\vec{k}|<k_{u}}
\zeta^{\dagger}_{\vec{k}}\zeta_{\vec{k}}
+\sum_{|\vec{k}|<k_{u}}\omega_{k}a_{\vec{k}}^{\dagger}a_{\vec{k}} \nonumber \\
 & & \mbox{}+\frac{e^{2}\rho}{2m}\sum_{|\vec{k}|<k_{u}}\frac{1}{2\omega_{k}}
[a_{\vec{k}}a_{-\vec{k}}+a_{\vec{k}}a_{\vec{k}}^{\dagger}
+a_{\vec{k}}^{\dagger}a_{\vec{k}}+a_{\vec{k}}^{\dagger}a_{-\vec{k}}^{\dagger}]
\end{eqnarray}
and
\begin{equation}
H_{int}^{(1)}=i\sum_{|\vec{k}|<k_{u}}g_{\vec{k}}
[a_{\vec{k}}(\zeta^{\dagger}_{\vec{k}}
-\zeta_{-\vec{k}})+a_{\vec{k}}^{\dagger}
(\zeta_{-\vec{k}}^{\dagger}-\zeta_{\vec{k}})].
\end{equation}

We first note that only the modes $\vec{k}$ and $-\vec{k}$ are coupled to each
other.  Therefore, we can diagonalize $H_{0}+H_{int}^{(1)}$ by considering modes
a pair at a time and adding up the results.  Let $H_{\vec{k}}$ be the part of
$H_{0}+H_{int}^{(1)}$ containing operators for the modes $\vec{k}$ and $-\vec{k}$.
It can be expressed as
\begin{equation}
H_{\vec{k}}=\nu^{\dagger}B\nu,
\end{equation}
where
\begin{equation}
\nu = \left(\begin{array}{c} a_{\vec{k}} \\ \zeta_{\vec{k}} \\ a_{-\vec{k}}^{\dagger} \\
\zeta_{-\vec{k}}^{\dagger} \end{array} \right)
\end{equation}
and
\begin{equation}
B=\left( \begin{array}{cccc} 
\omega_{k}+\frac{e^{2}\rho}{2m\omega_{k}} & -ig_{\vec{k}}
& \frac{e^{2}\rho}{2m\omega_{k}} & ig_{\vec{k}} \\ ig_{\vec{k}} & E_{0} & 
ig_{\vec{k}} & 0 \\ \frac{e^{2}\rho}{2m\omega_{k}} & -ig_{\vec{k}} &
\omega_{k}+\frac{e^{2}\rho}{2m\omega_{k}} & ig_{\vec{k}} \\ -ig_{\vec{k}} & 
0 & -ig_{\vec{k}} & E_{0} \end{array} \right)
\end{equation}
In arriving at Eq.(31) we have dropped all constants, 
such as those arising from
commutation relations, from the Hamiltonian, and used the fact that for
our choice of polarization, $g_{-\vec{k}}=g_{\vec{k}}$.

Following Hopfield we want to introduce operators which 
are linear combinations of 
the components of $\nu$.  We shall let $\alpha_{\vec{k}}$ 
and $\beta_{\vec{k}}$ be these new operators where we require
\begin{equation}
[\alpha_{\vec{k}},\alpha_{\vec{k}}^{\dagger}]=
[\beta_{\vec{k}},\beta_{\vec{k}}^{\dagger}]=1
\end{equation}
and
\begin{equation}
[\alpha_{\vec{k}},\beta_{\vec{k}}]=
[\alpha_{\vec{k}},\beta_{\vec{k}}^{\dagger}]=0.
\end{equation}
Defining the vector
\begin{equation}
v=\left( \begin{array}{c} 
\alpha_{\vec{k}} \\ \beta_{\vec{k}} \\ \alpha_{-\vec{k}}
^{\dagger} \\ \beta_{-\vec{k}}^{\dagger} \end{array} \right) ,
\end{equation}
we have that
\begin{equation}
v=M\nu,
\end{equation}
where $M$ is a 4x4 matrix yet to be determined.  
The commutation relations obeyed
by the elements of $v$ and $\nu$ imply that $M$ has the property that
\begin{equation}
M^{-1}=GM^{\dagger}G,
\end{equation}
where the matrix $G$ has the elements $G_{ij}=\delta_{ij}$ if $i,j=1,2$, and 
$G_{ij}=-\delta_{ij}$ if $i,j=3,4$, with all other matrix elements being   
zero.  From Eq. (38) we see immediately that M is not unitary.

Returning to $H_{\vec{k}}$ we have that
\begin{equation}
H_{\vec{k}}=v^{\dagger}(M^{-1})^{\dagger}BM^{-1}v,
\end{equation} 
which we want to be diagonal.  This implies that 
\begin{equation}
(M^{-1})^{\dagger}BM^{-1}=D
\end{equation}
where $D$ is a diagonal matrix, or, 
making use of Eq. (38) and the fact that $G^{2}=1$,
\begin{equation}
M(GB)M^{-1}=GD.
\end{equation}
Noting that if $D$ is diagonal so is $GD$ we see 
that $M$ must diagonalize the matrix $GB$.

We can reduce finding $M$ to finding the eigenvectors of $GB$.  
Let us denote the 
eigenvectors of $GB$ by $\eta_{i}$, where $i=1,\ldots,4$, and the unit vector 
whose only nonzero component is the ith by $c(i)$, 
i.\ e.\ $c(i)_{j}=\delta_{ij}$.
If $M$ satisfies the equation
\begin{equation}
M\eta_{i}=c(i),
\end{equation}
then we find that 
\begin{equation}
\langle c(i)|M(GB)M^{-1}|c(j)\rangle = \lambda_{i}\delta_{ij},
\end{equation}
where $\lambda_{i}$ is the eigenvalue of $GB$ corresponding to $\eta_{i}$.
Therefore, if $M$ satisfies Eq. (42) it will diagonalize $GB$.  
This equation also
gives us immediately an explicit representation for $M^{-1}$
\begin{equation}
M^{-1}=\sum_{i=1}^{4} |\eta_{i}\rangle \langle c(i)|.
\end{equation}
It is then possible to use Eq. (38) to find $M$
\begin{equation}
M=G(M^{-1})^{\dagger}G=G\sum_{i=1}^{4}|c(i)\rangle \langle \eta_{i}|G
\end{equation}
Finally, let us note that Eq. (38) imposes a 
condition on the eigenvectors $\eta_{i}$.
We have from this equation that 
\begin{equation}
(M^{-1})^{\dagger}GM^{-1}=G.
\end{equation}
Substitution of Eq.(42) into this result gives us that
\begin{equation}
\langle \eta_{j}|G|\eta_{i}\rangle = G_{ij}.
\end {equation}
For $i\neq j$ this condition is automatically satisfied because $\eta_{i}$ and
$\eta_{j}$ are eigenvalues of $GB$ corresponding to different eigenvalues. For
$i=j$ it can be imposed as a normalization condition.

The only task remaining is to find the eigenvectors 
and eigenvalues of $GB$.  Explicit
expressions are given in Appendix B along with those 
for the matrix elements of $M$.
Those results along with Eqs. (31) and (37) imply that
\begin{equation}
H_{\vec{k}}=E_{1}(k)\alpha^{\dagger}_{\vec{k}}\alpha_{\vec{k}} 
+E_{2}(k)\beta^{\dagger}
_{\vec{k}}\beta_{\vec{k}}+E_{1}(k)\alpha^{\dagger}_{-\vec{k}}\alpha_{-\vec{k}} 
+E_{2}(k)\beta^{\dagger}_{-\vec{k}}\beta_{-\vec{k}},
\end{equation}
where
\begin{eqnarray}
E_{1}(k) & = & \frac{1}{\sqrt{2}}( [E_{0}^{2}+\omega_{k}(\omega_{k}+2C_{0})] 
\nonumber\\
 & & \mbox{} +[ [E_{0}^{2}-\omega_{k}(\omega_{k}+2C_{0})]^{2}
+16E_{0}\omega_{k}g_{\vec{k}}^{2}]^{1/2})^{1/2} 
\end{eqnarray}
\begin{eqnarray}
E_{2}(k) & = &\frac{1}{\sqrt{2}}( [E_{0}^{2}+\omega_{k}(\omega_{k}+2C_{0})] 
\nonumber \\
 & & \mbox{} -[ [E_{0}^{2}-\omega_{k}(\omega_{k}+2C_{0})]^{2}
+16E_{0}\omega_{k}g_{\vec{k}}^{2}]^{1/2})^{1/2},
\end{eqnarray}
with $C_{0}=(e^{2}\rho)/(2m\omega_{k})$.  All of the square roots in the 
above equations are taken to be positive.  
The energies $E_{1}(k)$ and $E_{2}(k)$
are just the two branches of the polariton energy curve.  For large values of
$\omega_{k}$ we find that 
$E_{1}(k)\rightarrow \omega_{k}$ and $E_{2}(k)\rightarrow
 E_{0}$.  As $\omega_{k}\rightarrow 0$ we find that $E_{1}(k)$ goes to a value
slightly larger than $E_{0}$
\begin{equation}
E_{1}(k)\rightarrow E_{0}+2\frac{\omega_{k}g_{\vec{k}}^{2}}{E_{0}^{2}},
\end{equation}
and $E_{2}(k)$ goes to
\begin{equation}
E_{2}(k)= \left( 2\omega_{k}C_{0}-\frac{4\omega g_{\vec{k}}^{2}}{E_{0}}
+\omega_{k}^{2} \right)^{1/2}.
\end{equation}
It should be noted that $\omega_{k}C_{0}$ and $\omega_{k}g_{\vec{k}}^{2}$
are independent of $\omega_{k}$, and that $\omega_{k}g_{\vec{k}}^{2}
<< E_{0}^{3}$ and $\omega_{k}C_{0}<< E_{0}^{2}$.
In Hopfield's analysis the small constant term inside the 
parentheses vanished due
to an atomic sum rule.  This tells us that keeping only 
one excited atomic level
is not a good approximation near $\omega_{k}=0$.  
Because we are interested in optical
phenomena this does not present a problem.

We can finally express $H_{0}+H_{int}^{(1)}$ 
in terms of the polariton operators.
Dropping constant terms we have
\begin{equation}
H_{0}+H_{int}^{(1)}=\sum_{|\vec{k}|<k_{u}}
(E_{1}(k)\alpha^{\dagger}_{\vec{k}}\alpha_{\vec{k}}
+E_{2}(k)\beta_{\vec{k}}^{\dagger}\beta_{\vec{k}}).
\end{equation}
The effect of the medium appears in two ways in this Hamiltonian.  First, the 
operators are mixed matter-field operators, 
i.\ e.\ polariton, not photon, operators.
Second, the effects of dispersion appear through the energies $E_{1}(k)$ and
$E_{2}(k)$ which are not of photon form.  Therefore, by including the matter
degrees of freedom in the theory dispersion 
emerges naturally, and we avoid the 
problems of trying to quantize a theory which is nonlocal in time.

\section{Nonlinear Interaction}
We now turn our attention to the nonlinear 
part of the Hamiltonian, $H_{int}^{(2)}$,
which can be interpreted as describing an 
interaction between polaritons.  With
the interaction expressed as in Eq. (25), however, this interpretation is not 
obvious.  In order to bring it out we begin by expressing it in terms of the
operators $\zeta_{\vec{k}}$
\begin{eqnarray}
H_{int}^{(2)} & = & -\frac{i}{2\rho V}\sum_{|\vec{k}|<k_{u}}
\ldots \sum_{|\vec{k}_{3}|<k_{u}}
g_{\vec{k}}(\delta_{\vec{k}+\vec{k}_{3},\vec{k}_{1}+\vec{k}_{2}}a_{\vec{k}}
\zeta^{\dagger}_{\vec{k}_{1}}\zeta^{\dagger}_{\vec{k}_{2}}\zeta_{\vec{k}_{3}} 
\nonumber\\
 & & \mbox{}-\delta_{\vec{k}+\vec{k}_{2}+\vec{k}_{3},\vec{k}_{1}}a_{\vec{k}}
\zeta^{\dagger}_{\vec{k}_{1}}\zeta_{\vec{k}_{2}}\zeta_{\vec{k}_{3}}
+\delta_{\vec{k}+\vec{k}_{2}+\vec{k}_{3},\vec{k}_{1}}a_{\vec{k}}^{\dagger}
\zeta^{\dagger}_{\vec{k}_{2}}\zeta_{\vec{k}_{3}}^{\dagger}\zeta_{\vec{k}_{1}} 
\nonumber\\
 & & \mbox{}-\delta_{\vec{k}+\vec{k}_{3},\vec{k}_{1}+\vec{k}_{2}}
a_{\vec{k}}^{\dagger}
\zeta^{\dagger}_{\vec{k}_{3}}\zeta_{\vec{k}_{2}}\zeta_{\vec{k}_{1}}).
\end{eqnarray}
The next step is to express the photon and atomic 
operators in terms of polariton
operators.  This leads to a great many terms so it 
is perhaps best to consider a
specific physical process and then to select the terms which are relevant to it.

Let us first consider the situation when a single polariton mode, for example, 
the one corresponding to the operator $\alpha_{\vec{k}_{0}}$ in the regime 
where $\omega_{k_{0}}>E_{0}$, is highly excited.  The 
dominant terms in $H_{int}^{(2)}$, at least 
for times which are not too long, will
be those in which each of the four operators 
refers to the excited mode.  The other
terms will have a smaller effect on the 
time evolution because they contain at least
one operator for a mode which is initially in the vacuum state.  
Examining the operator transformations in Eq. (37) we see that if 
we only keep the polariton operators $\alpha_{\vec{k}_{0}}$ 
and $\alpha_{\vec{k}_{0}}^{\dagger}$, then
\begin{eqnarray}
a_{k_{0}}\rightarrow A_{11}(\vec{k}_{0})\alpha_{\vec{k}_{0}}&\hspace{2cm}&
a_{-k_{0}}\rightarrow A_{31}^{\ast}(\vec{k}_{0})
\alpha_{\vec{k}_{0}}^{\dagger} \nonumber\\
\zeta_{k_{0}}\rightarrow A_{21}(\vec{k}_{0})\alpha_{\vec{k}_{0}} 
& \hspace{2cm} &
\zeta_{-k_{0}}\rightarrow A_{41}^{\ast}(\vec{k}_{0})
\alpha_{\vec{k}_{0}}^{\dagger}.
\end{eqnarray}
where have set $A=M^{-1}$.  In the regime we have chosen, $\omega_{k_{0}}
>E_{0}$ on the $\alpha$ branch, we find that $A_{11}(\vec{k}_{0})$ is 
of order one, $A_{21}(\vec{k}_{0})$ and $A_{41}(\vec{k}_{0})$ are of
order $g_{\vec{k}_{0}}/E_{0}$, and $A_{31}(\vec{k}_{0})$ is of order
$C_{0}/E_{0}$ which is considerably smaller than $g_{\vec{k}_{0}}/E_{0}$.
Therefore, we shall drop terms containing $A_{31}(\vec{k}_{0})$.  Keeping
this in mind and making the substitutions indicated in Eq. (55) in
Eq. (54) we find
\begin{eqnarray}
H_{int}^{(dom)} &= &-\frac{ig_{\vec{k}_{0}}}{2\rho V}A_{11}[2|A_{21}|^{2}
A_{21}^{\ast}(\alpha_{\vec{k}_{0}}^{\dagger})^{2}
(\alpha_{\vec{k}_{0}})^{2} \nonumber\\
 & &\mbox{}+2|A_{41}|^{2}A_{21}^{\ast}\alpha_{\vec{k}_{0}}^{\dagger}
\alpha_{\vec{k}_{0}}\alpha_{\vec{k}_{0}}^{\dagger}\alpha_{\vec{k}_{0}}
+(|A_{41}|^{2}A_{21}^{\ast}-|A_{21}|^{2}A_{41}^{\ast})
\alpha_{\vec{k}_{0}}(\alpha_{\vec{k}_{0}}^{\dagger})^{2}\alpha_{\vec{k}_{0}}
\nonumber\\
 & &\mbox{}+(|A_{21}|^{2}A_{41}-|A_{41}|^{2}A_{21})
\alpha_{\vec{k}_{0}}^{\dagger}(\alpha_{\vec{k}_{0}})^{2}
\alpha_{\vec{k}_{0}}^{\dagger} + 2|A_{21}|^{2}A_{41}\alpha_{\vec{k}_{0}}
\alpha_{\vec{k}_{0}}^{\dagger}\alpha_{\vec{k}_{0}}
\alpha_{\vec{k}_{0}}^{\dagger} \nonumber\\
 & & \mbox{}+2|A_{41}|^{2}A_{41}(\alpha_{\vec{k}_{0}})^{2}
(\alpha_{\vec{k}_{0}}^{\dagger})^{2}],
\end{eqnarray}
where all of the matrix elements are evaluated at $\vec{k}_{0}$, and
we have made use of the fact that $A_{11}(\vec{k}_{0})$ is real and both
$A_{21}(\vec{k}_{0})$ and $A_{41}(\vec{k}_{0})$ are imaginary.

Let us note several things about this expression.  First, it looks
similar to what we would expect an interaction which describes
self-phase modulation to look like.  Therefore, by going to a description
in terms of polariton operators we have recovered a familiar nonlinear 
optical interaction.  Second, it is not normally ordered.  If we use
commutators to bring it into normal order we will pick up terms 
proportional to $\alpha_{\vec{k}_{0}}^{\dagger}\alpha_{\vec{k}_{0}}$. 
These represent small shifts to the polariton frequency and can be 
neglected in most applications.  Neglecting them gives us
\begin{equation}
H_{int}^{(dom)} \cong \chi (\alpha_{\vec{k}_{0}}^{\dagger})^{2}
\alpha_{\vec{k}_{0}}^{2}
\end{equation}
where
\begin{equation}
\chi = -\frac{ig_{\vec{k}_{0}}}{2\rho V}A_{11}[2(|A_{21}|^{2}+2|A_{41}|^{2})
A_{21}^{\ast}+2(|A_{41}|^{2}+2|A_{21}|^{2})A_{41}].
\end{equation}

The effect of self-phase modulation on a single mode is often described by
the Hamiltonian
\begin{equation}
H_{int}=\lambda (a^{\dagger})^{2}a^{2}.
\end{equation}
which is superficially similar to Eq. (57).  Hoever, in most treatments
the operators appearing in Eq. (59) are assumed to be photon creation
and annihilation operators.  As is shown by Eq. (57), they should be
interpreted as polariton operators instead.

Now let us consider a more complicated situation.  
Suppose we initially have
a pulse which is made up of modes on the $\alpha $ 
branch with wave vectors near $\vec{k}_{0}$.  In particular, let us assume
that all of the wave vectors of the modes present in the pulse lie in
a small region $S$ about $\vec{k}_{0}$.  Keeping only terms in $H_{int}^{(2)}$ 
which contain four excited modes (after transforming to polariton operators)
we obtain a rather complicated interaction.  Each term in it contains two
creation and two annihilation operators, and four elements of the matrix
$A(\vec{k})$, each evaluated at a different wave vector.  The interaction
simplifies considerably if we make two approximations.  First, we ignore
operator ordering, which, as we saw, is tantamount to neglecting small
frequency shifts.  Second, we approximate each matrix element of $A$ by
its value at $\vec{k}_{0}$.  Because the spread in wave vectors is small
this is a good approximation.  With these approximations we find
\begin{equation}
H_{int}^{(dom)} \cong \chi\sum_{\vec{k}\in S}\sum_{\vec{k}_{1}\in S}
\sum_{\vec{k}_{2}\in S}\sum_{\vec{k}_{3}\in S}\delta_{\vec{k}+\vec{k}_{1},
\vec{k}_{2}+\vec{k}_{3}}\alpha_{\vec{k}}^{\dagger}
\alpha_{\vec{k}_{1}}^{\dagger}\alpha_{\vec{k}_{2}}\alpha_{\vec{k}_{3}}.
\end{equation}

This situation is often treated by using a phenomenological Hamiltonian
which describes pulse propagation in a $\chi^{(3)}$ medium [5].  The
Hamiltonian is
\begin{equation}
H=\int dx\left[ \frac{\partial \phi^{\dagger}}{\partial x}\frac{\partial \phi}
{\partial x} +c(\phi^{\dagger})^{2}\phi^{2} \right],
\end{equation}
where the field $\phi (x,t)$ is the field envelope of the pulse in a frame
moving at the group velocity of the pulse, and it obeys the
commutation relations
\begin{equation}
[\phi (x,t),\phi^{\dagger}(x^{\prime},t)]=\delta (x-x^{\prime}).
\end{equation}
The equation of motion for the field operator $\phi (x,t)$ resulting from this
theory is the nonlinear Schroedinger equation.  It can also be
expressed in terms of creation and annihilation operators.  Defining
the annhilation operator
\begin{equation}
a(\beta,t)=\frac{1}{\sqrt{2\pi}}\int dx e^{i\beta x}\phi (x,t),
\end{equation}
we find for the equation of motion [5]
\begin{equation}
i\frac{\partial a(\beta,t)}{\partial t}=\beta^{2}a(\beta,t) +2c\int d\beta_{1}
\int d\beta_{2} a^{\dagger}(\beta_{1},t)a(\beta_{2},t)
a(\beta +\beta_{1}-\beta_{2},t).
\end{equation}

It should be pointed out that the $t$ in these equations 
is not really time and the
$x$ is not really space.  The above formalism is derived by an analogy to the
classical theory which describes the propagation of a pulse in a nonlinear
dispersive medium.  The pulse is assumed to be propagating in the $z$ direction.
The variable $x$ is $v_{g}t-z$, where $v_{g}$ is the group velocity, and $t$ is
proportional to $z$.  This means that what look like equal-time commutation
relations in Eq. (62) are actually equal-space commutation relations.  
The question of when a theory using equal-space commutation relations 
gives the same results as a canonically quantized theory has not 
been fully answered.
Deutsch has done a preliminary investigation and found that for a linear
theory they are equivalent [12].  However, for a nonlinear theory he found
indications that field correlation functions in which the fields are
evaluated at different spatial points will not be the same in the two theories. 
Because of the different commutation relations, it is rather difficult to 
directly compare Eq. (64) to
an equation of motion derived from our interacting polariton theory.

Instead we shall compare it to one due to Carter and Drummond who derived it by
quantizing the macroscopic theory [2].  
Their theory describes a pulse, consisting
of modes with wave numbers centered about $k_{c}$ 
and frequencies centered about
$\omega_{c}$, propagating in a medium of length $L$ 
in the positive $z$ direction.
The basic field in this theory is 
\begin{equation}
\Psi (z,t) = e^{-ik_{c}z+i\omega_{c}t}\frac{1}{\sqrt{L}}
\sum_{k}e^{ikz}a_{k}(t),
\end{equation}
which, because of the initial exponential factor, 
is slowly varying in space and
time.  The annihilation operators are defined in terms, not of the vector
potnetial and the electric field, but in terms of the dual potential and the
displacement field [2,13].  They, therefore, 
implicitly contain matter degrees of
freedom and are, in a sense, polariton operators.  The field $\Psi(z,t)$ obeys
the equal-time comutation relations
\begin{equation}
[\Psi(z,t), \Psi^{\dagger}(z^{\prime},t)]= \delta (z-z^{\prime}).
\end{equation}

Like ours, the Carter-Drummond theory describes the pulse in the lab frame,
i.\ e.\ the frame in which the medium is at rest.  
This removes one of the difficulties 
we had in trying to compare our theory 
with the phenomenological nonlinear-Schroedinger equation 
theory which is formulated in a moving frame.  Carter and Drummond also
reformulate their theory in a moving frame in order to make a comparison to the 
phenomenological theory, but we shall use their original lab frame results.  In
this frame their Hamiltonian is
\begin{eqnarray}
H & = & \frac{1}{2} \int_{0}^{L}dz\left[iv \left(\frac{\partial \Psi^{\dagger}}
{\partial z}\Psi - \Psi^{\dagger}\frac{\partial \Psi}{\partial z}\right) 
\right. \nonumber \\
 & & \left . +\omega'' \frac{\partial \Psi^{\dagger}}{\partial z}
\frac{\partial \Psi}
{\partial z} - v^{2}\chi^{E}(\Psi^{\dagger})^{2}\Psi^{2}\right],
\end{eqnarray}
where $v$ is the group velocity of the pulse, 
$\chi^{E}$ is proportional to the 
third order nonlinear susceptibility, and $\omega''$ is the second derivative
of the frequency with respect to the wave number evaluated at $k_{c}$.  
The Hamiltonian
and the commutation relations give, for the equation 
of motion for the operator $a_{k}(t)$,
\begin{eqnarray}
i\frac{da_{k}}{dt} & = & [\omega_{c}+v(k-k_{c})+
\frac{1}{2}\omega''(k-k_{c})^{2}]a_{k} \nonumber \\
 & & \mbox{}-v^{2}\chi^{E}\frac{1}{L}\sum_{k_{1}}
\sum_{k_{2}}a^{\dagger}_{k_{1}}a_{k_{2}}a_{k+k_{1}-k_{2}}.
\end{eqnarray}
It is this equation which we wish to compare to 
the corresponding equation derived
from our nonlinear polariton theory.

Let us find the equation of motion for $\alpha_{\vec{k}}(t)$.
We shall assume that we have a pulse made up of 
wave vectors near $\vec{k}_{0}$ so
that the interaction is Eq. (60) is appropriate.  Because all wave vectors
in the pulse are close to $\vec{k}_{0}$, we shall
expand the polariton energy $E_{1}(\vec{k})$ about $\vec{k}_{0}$
\begin{eqnarray}
E_{1}(\vec{k}) & = & E_{1}(\vec{k}_{0})+(\delta\vec{k}\cdot\hat{k}_{0})
\frac{dE_{1}}{dk}
+\frac{1}{2k_{0}}(\delta\vec{k}^{2}-(\delta\vec{k}\cdot\hat{k}_{0})^{2})
\frac{dE_{1}}{dk} \nonumber \\
 & & \mbox{}+\frac{1}{2}(\delta\vec{k}\cdot\hat{k}_{0})^{2}
\frac{d^{2}E_{1}}{dk^{2}},
\end{eqnarray}
where $\delta\vec{k}=\vec{k}-\vec{k}_{0}$, 
$\hat{k}_{0}=\vec{k}_{0}/k_{0}$, and
all derivatives are evaluated at $k_{0}=|\vec{k}_{0}|$.
Setting $v=dE_{1}/dk$, we find 
\begin{eqnarray}
i\frac{d{\alpha}_{\vec{k}}}{dt} & \cong & [E_{1}(k_{0})+v\hat{k_{0}}\cdot
\delta\vec{k}+\frac{1}{2k_{0}}(\delta\vec{k}^{2}
-(\delta\vec{k}\cdot\hat{k}_{0})^{2})v +
\frac{1}{2}(\delta\vec{k}\cdot\hat{k}_{0})^{2}
\frac{d^{2}E_{1}}{dk^{2}}]{\alpha}_{\vec{k}} \nonumber \\
 & & \mbox{}+2\chi
\sum_{\vec{k}_{1}\in S}\sum_{\vec{k}_{2}\in S}{\alpha}^{\dagger}_{\vec{k}_{1}}
{\alpha}_{\vec{k}_{2}}{\alpha}_{\vec{k}+\vec{k}_{1}-\vec{k}_{2}}.
\end{eqnarray}
If the pulse is one dimensional, i.\ e.\ if $\delta\vec{k}$ 
is always parallel to
$\hat{k}_{0}$, then Eqs. (68) and (70) have the same form.  
Thus we recover from the 
microscopic theory an equation of motion of the same form as that which arises
from the quantized macroscopic theory.  Note that extending the treatment here
to broadband pulses is straightforward for microscopic theory; one simply does
not expand the polariton energy as a function of $\vec{k}$.  This extension is 
more complicated for the quantized macroscopic theory.

We can also examine the more complicated situation 
in which two modes are initially
highly excited.  Let us look at two cases, one when 
the modes are counterpropagating
and one when they are not.  The counterpropagating case 
is the more complicated
of the two so we shall consider it second.

Suppose that the two polariton modes on the $\alpha$ branch with wave vectors
$\vec{k}_{a}$ and $\vec{k}_{b}$ are initially highly excited.  
Let us assume that 
$\vec{k}_{a}$ and $\vec{k}_{b}$ are both in the $x$-$y$ plane, and that both
modes are polarized in the $z$ direction.  
We shall also assume that $\vec{k}_{a} \neq -\vec{k}_{b}$.  
Keeping only the two excited modes, and making the substitutions in
Eq. (55) for each mode, the Hamiltonian becomes
\begin{equation}
H_{int}^{(dom)}\cong \chi_{a}(\alpha_{\vec{k}_{a}}^{\dagger})^{2}
\alpha_{\vec{k}_{a}}^{2}+\chi_{b}(\alpha_{\vec{k}_{b}}^{\dagger})^{2}
\alpha_{\vec{k}_{b}}^{2}+\chi_{ab}\alpha_{\vec{k}_{a}}^{\dagger}
\alpha_{\vec{k}_{b}}^{\dagger}\alpha_{\vec{k}_{a}}\alpha_{\vec{k}_{b}}. 
\end{equation}
Here $\chi_{a}$ is simply $\chi$ with all of the matrix elements evaluated
at $\vec{k}_{a}$, with a similar definition for $\chi_{b}$, and $\chi_{ab}$
is given by
\begin{equation}
\chi_{ab}=-\frac{i}{\rho V}(F-F^{\ast}),
\end{equation}
where
\begin{eqnarray}
F & = & A_{11}^{(a)}g_{\vec{k}_{a}}[(|A_{21}^{(b)}|^{2}+|A_{41}^{(b)}|^{2})
(A_{21}^{(a)\ast}+A_{41}^{(a)})+A_{21}^{(b)\ast}A_{41}^{(b)}(A_{41}^{(a)\ast}
+A_{21}^{(a)})] \nonumber\\
 & & A_{11}^{(b)}g_{\vec{k}_{b}}[(|A_{21}^{(a)}|^{2}+|A_{41}^{(a)}|^{2})
(A_{21}^{(b)\ast}+A_{41}^{(b)}) \nonumber \\
 & & \mbox{}+A_{21}^{(a)\ast}A_{41}^{(a)}(A_{41}^{(b)\ast}
+A_{21}^{(b)})]
\end{eqnarray}
In the above equation the superscript on the matrix element of $A$ 
indicates whether it is evaluated at $\vec{k}_{a}$ or $\vec{k}_{b}$.
In Eq. (71) one has the usual terms which describe 
cross- and self-phase modulation of
the two modes.  It should be noted that if $|\vec{k}_{a}|=|\vec{k}_{b}|$, then
with the above stated conditions on $\vec{k}_{a}$ and $\vec{k}_{b}$, and the
polarizations, we have that $A(\vec{k}_{a})=A(\vec{k}_{b})$.  
This in turn implies that $\chi_{ab}=4\chi_{a}=4\chi_{b}$, and
that the interaction in Eq. (73) can be 
derived from the kind of interaction appearing
in the Hamiltonian in Eq. (61) or Eq. (67).  
If $|\vec{k}_{a}|\neq |\vec{k}_{b}|$,
then this is no longer the case, but will be approximately true if 
$|\vec{k}_{a}|-|\vec{k}_{b}|$ is small and the dependence of $A(\vec{k})$
on $|\vec{k}|$ is weak.

Now let us look at the case of two counterpropagating modes.  We shall assume
that two modes on the $\alpha$ branch with wave vectors $\vec{k}_{0}$ and
$-\vec{k}_{0}$, and with the same polarization, 
are initially highly excited.  We
again want to keep only these two modes in our Hamiltonian, but the situation
is now considerably more complicated 
than in our previous cases.  This is because each of the matter or field
operators will now be replaced by a sum of two polariton operators, e.\ g.\ 
\begin{eqnarray}
a_{\vec{k}_{0}} & \rightarrow & A_{11}(\vec{k}_{0})\alpha_{\vec{k}_{0}}
+A_{13}(\vec{k}_{0})\alpha_{-\vec{k}_{0}}^{\dagger} \nonumber \\
\zeta_{\vec{k}_{0}} & \rightarrow & A_{21}(\vec{k}_{0})\alpha_{\vec{k}_{0}}
+A_{23}(\vec{k}_{0})\alpha_{-\vec{k}_{0}}^{\dagger},
\end{eqnarray}
where contributions from initially unpopulated modes have been dropped.  We
find that
\begin{eqnarray}
H_{int}^{(2)} & \cong & \frac{ig_{\vec{k}_{0}}}{2\rho V}A_{11}(A_{21}-A_{23})
\{6A_{21}A_{23}[(\alpha_{\vec{k}_{0}}^{\dagger}
\alpha_{-\vec{k}_{0}}^{\dagger})^{2}
+(\alpha_{\vec{k}_{0}}\alpha_{-\vec{k}_{0}})^{2}] \nonumber \\
 & & \mbox{}+3(A_{21}-A_{23})^{2}
[(\alpha_{\vec{k}_{0}}^{\dagger})^{2}\alpha_{-\vec{k}_{0}}^{\dagger}
\alpha_{\vec{k}_{0}}
+(\alpha_{-\vec{k}_{0}}^{\dagger})^{2}\alpha_{\vec{k}_{0}}^{\dagger}
\alpha_{-\vec{k}_{0}}+\alpha_{-\vec{k}_{0}}^{\dagger}\alpha_{-\vec{k}_{0}}^{2}
\alpha_{\vec{k}_{0}} \nonumber \\
 & & \mbox{}+\alpha_{\vec{k}_{0}}^{\dagger}\alpha_{\vec{k}_{0}}^{2}
\alpha_{-\vec{k}_{0}}]-2(A_{21}^{2}-A_{21}A_{23}+A_{23}^{2})
[(\alpha_{\vec{k}_{0}}^{\dagger})^{2}\alpha_{\vec{k}_{0}}^{2} \nonumber \\
 & & \mbox{}+(\alpha_{-\vec{k}_{0}}^{\dagger})^{2}\alpha_{-\vec{k}_{0}}^{2}
+4 \alpha_{\vec{k}_{0}}^{\dagger}\alpha_{-\vec{k}_{0}}^{\dagger}
\alpha_{\vec{k}_{0}}\alpha_{-\vec{k}_{0}}\},
\end{eqnarray}
where all of the matrix elements are evaluated at $\vec{k}_{0}$.  
In deriving Eq. (75)
we have made use of the fact that $A_{11}$ and $A_{13}$ 
are real, that $A_{21}$
and $A_{23}$ are imaginary (see Appendix B), and that $A_{13}$ can be
neglected in comparison to the other three.
We have also dropped terms due to 
commutators.  These are similar in form to the 
terms in Eq. (75) except that they
contain only two $\alpha_{\vec{k}}$ operators instead of four.  They are,
therefore, smaller than the terms we have kept by a factor of the order of the
number of photons in the $\alpha_{\vec{k}_{0}}$ and $\alpha_{-\vec{k}_{0}}$
modes.  If we were to examine the population of initially unpopulated modes
by the two counterpropagating beams, these terms should be kept.  In particular,
these terms would play a role in the four-wave 
mixing process which is responsible for phase conjugation and squeezing.

The interaction in Eq. (75) cannot be derived 
from an effective Hamiltonian like that 
in Eq. (61).  Eq. (75) contains terms with 
unequal numbers of creation and annihilation
operators while Eq. (61) does not.  This means that 
Eq. (61) cannot be used to treat
counterpropagating pulses in a nonlinear medium, but that a more complicated
Hamiltonian, such as that in Eq. (75), must be used.

\section{Conclusion}
The justification of the Hamiltonians which are used in the quantum theory of
nonlinear optics is an important part of placing this theory on a firmer basis.
We have presented a derivation of an effective Hamiltonian for the interaction
of light and a medium consisting of
two-level atoms which contains a number of the standard nonlinear optical
interactions.  This is accomplished 
by expanding the atomic operators in terms of boson
creation and annihilation operators and 
diagonalizing the part of the Hamiltonian
which describes linear interactions.  The result is a theory of interacting 
polaritons which is a
nonlinear extension of Hopfield's theory.  Because the dispersion relation
for polaritons is different from that of photons, the effects of dispersion are
automatically included in this theory.

The theory presented here provides further justification for the
Hamiltonians which emerge from the quantized macroscopic theory.  In addition,
it can be used directly to provide effective Hamiltonians
for more complicated situations than
those which have so far been considered.  We saw an example of this in the case
of counterpropagating beams.  The microscopic theory serves as a useful
counterpart to the quantized macroscopic theory in providing a description of
quantized fields in nonlinear dielectric media.

\section*{Acknowledgments}
This research was supported by the National Science Foundation under grant
number PHY- 9403601 and a grant from the PSC-CUNY Research Award Program.
One of us (M.\ H.\,) would like to thank Peter Drummond for useful
conversations and for providing to us a copy of his review article with S. 
Carter before publication.

\section*{Appendix A}
Here we examine the ground state of the Hamiltonian in the lowest order of the
semiclassical expansion.  The spin operators are replaced 
by the c-number quantities
\begin{eqnarray}
S_{l}^{(3)} & \rightarrow & s\cos \theta_{l} \nonumber \\
S_{l}^{(+)} & \rightarrow & se^{i\phi_{l}}\sin\theta_{l} \nonumber \\
S_{l}^{(-)} & \rightarrow & se^{-i\phi_{l}}\sin\theta_{l},
\end{eqnarray}
giving us the Hamiltonian
\begin{eqnarray}
H_{sc} & = & \sum_{l=1}^{N_{b}}E_{0}s\cos\theta_{l}
+\sum_{|\vec{k}|<k_{u},\lambda}
\omega_{k}a_{\vec{k}\lambda}^{\dagger}a_{\vec{k}\lambda} \nonumber \\
 & & \mbox{}-\sum_{|\vec{k}|<k_{u},\lambda}
\sum_{l=1}^{N_{b}}(a_{\vec{k}\lambda}
e^{i\vec{k}\cdot\vec{r}_{l}}+a_{\vec{k}\lambda}^{\dagger}
e^{-i\vec{k}\cdot\vec{r}_{l}})D_{l\vec{k}\lambda} \nonumber \\
 & & \mbox{}+\frac{e^{2}\rho}{2m}\sum_{|\vec{k}|<k_{u},\lambda}
\frac{1}{2\omega_{\vec{k}}}
[(-1)^{\lambda}a_{\vec{k}\lambda}a_{-\vec{k}\lambda}+a_{\vec{k}\lambda}
a_{\vec{k}\lambda}^{\dagger} \nonumber \\
 & & \mbox{}+a_{\vec{k}\lambda}^{\dagger}a_{\vec{k}\lambda}+(-1)^{\lambda}
a_{\vec{k}\lambda}^{\dagger}a_{-\vec{k}\lambda}^{\dagger}],
\end{eqnarray}
where
\begin{equation}
D_{l\vec{k}\lambda}=\frac{2s}{\sqrt{2\omega_{k}V}}\sin\theta_{l}\sin\phi_{l}
\mu_{\lambda}(\vec{k}).
\end{equation}
We have assumed that $\mu_{\lambda}(\vec{k})$ is real which implies that
$D_{l\vec{k}\lambda}$ is also real.  
In principle, we should diagonalize this Hamiltonian
which would allow us to determine the values of $\theta_{l}$ and $\phi_{l}$ as
well as the field in the ground state to lowest order.  In practice, we
shall be able to obtain what we need after completing part of the procedure,
and the rest of the diagonalization can be completed when additional
terms from $H_{0}$ and $H_{int}^{(1)}$ are included.

We begin by eliminating the terms linear in the field operators.  This can be
accomplished by shifting the creation and annihilation operators, i. e. by
setting
\begin{equation}
a_{\vec{k}\lambda}=b_{\vec{k}\lambda}+z_{\vec{k}\lambda},
\end{equation}
where $z_{\vec{k}\lambda}$ is a c number, and
\begin{equation}
b_{\vec{k}\lambda}=D_{\vec{k}\lambda}(-z_{\vec{k}\lambda})a_{\vec{k}\lambda}
D_{\vec{k}\lambda}(-z_{\vec{k}\lambda})^{-1}.
\end{equation}
The displacement operator $D_{\vec{k}\lambda}(z)$ is equal to $\exp(z
a_{\vec{k}\lambda}^{\dagger}-z^{\ast}a_{\vec{k}\lambda})$.  We now substitute
Eq. (79) into $H_{sc}$ and collect the terms linear in $b_{\vec{k}\lambda}$ and
$b_{\vec{k}\lambda}^{\dagger}$.  These are
\begin{eqnarray}
 & & \sum_{|\vec{k}|<k_{u},\lambda} \{ \omega_{k}(z_{\vec{k}\lambda}^{\ast}
b_{\vec{k}\lambda}+z_{\vec{k}\lambda}b_{\vec{k}\lambda}^{\dagger})-
L_{\vec{k}\lambda}b_{\vec{k}\lambda}-L_{\vec{k}\lambda}^{\ast}
b_{\vec{k}\lambda}^{\dagger} \nonumber \\
 & &\mbox{}+ \frac{e^{2}\rho}{4m\omega_{k}}[(-1)^{\lambda}(z_{\vec{k}\lambda}
b_{-\vec{k}\lambda}+z_{-\vec{k}\lambda}b_{\vec{k}\lambda})
+2(z_{\vec{k}\lambda}^{\ast}
b_{\vec{k}\lambda}+z_{\vec{k}\lambda}b_{\vec{k}\lambda}^{\dagger}) \nonumber \\
 & &\mbox{}+(-1)^{\lambda}(z_{\vec{k}\lambda}^{\ast}
b_{-\vec{k}\lambda}^{\dagger}
+z_{-\vec{k}\lambda}^{\ast}b_{\vec{k}\lambda}^{\dagger})]\},
\end{eqnarray}
where
\begin{equation}
L_{\vec{k}\lambda}=\sum_{l=1}^{N_{b}}e^{i\vec{k}\cdot\vec{r}_{l}}
D_{l\vec{k}\lambda}.
\end{equation}
By grouping the terms for $\vec{k}$ and $-\vec{k}$ 
together we find that the expression in 
Eq. (81) will vanish if
\begin{equation}
\omega_{k}z_{\vec{k}\lambda}\left(1
+\frac{e^{2}\rho}{2m\omega_{k}^{2}+e^{2}\rho}
\right) = L_{\vec{k}\lambda}-(-1)^{\lambda}
\frac{e^{2}\rho L_{\vec{k}\lambda}^{\ast}}
{2m\omega_{k}^{2}+e^{2}\rho}
\end{equation}
Making use of the fact that $L_{-\vec{k}\lambda}^{\ast}=(-1)^{\lambda}
L_{\vec{k}\lambda}$ this simplifies to
\begin{equation}
z_{\vec{k}\lambda}=\frac{mL_{\vec{k}\lambda}^{\ast}\omega_{k}}{m\omega_{k}^{2}
+e^{2}\rho}.
\end{equation}
Substitution of this result back into $H_{sc}$ gives us
\begin{eqnarray}
H_{sc}&=& \sum_{l=1}^{N_{b}}E_{0}s\cos\theta_{l}-\varepsilon_{0}+\sum_{\vec{k}
<k_{u},\lambda}\omega_{k}b_{\vec{k}\lambda}^{\dagger}
b_{\vec{k}\lambda}\nonumber \\
 & &\mbox{}+\frac{e^{2}\rho}{2m}
\sum_{\vec{k}<k_{u},\lambda}\frac{1}{2\omega_{k}}
[(-1)^{\lambda}b_{\vec{k}\lambda}b_{-\vec{k}\lambda}
+b_{\vec{k}\lambda}b_{\vec{k}\lambda}^{\dagger} \nonumber \\
 & & \mbox{}+b_{\vec{k}\lambda}^{\dagger}b_{\vec{k}\lambda}+(-1)^{\lambda}
b_{\vec{k}\lambda}^{\dagger}b_{-\vec{k}\lambda}^{\dagger}],
\end{eqnarray}
where 
\begin{equation}
\varepsilon_{0}=\sum_{\vec{k}<k_{u},\lambda}\frac{m|L_{\vec{k}\lambda}|^{2}\omega_{k}}
{m\omega_{k}^{2}+e^{2}\rho}.
\end{equation}

We can find the semiclassical ground state 
spin configuration by minimizing the 
sum of the first two terms in Eq. (85) 
with respect to $\phi_{l}$ and $\theta_{l}$.
That is, we minimize the expression
\begin{equation}
\sum_{l=1}^{N_{b}}E_{0}s\cos\theta_{l}-\varepsilon_{0}.
\end{equation}
We can obtain an estimate of what the 
minimum configuration will be by examining
the size of $\varepsilon_{0}$.  
If $\varepsilon_{0} <<N_{B}sE_{0}$, then the first 
term will be the dominant one, and the minimum will occur approximately at
$\theta_{l}=\pi$, for all $l$. This we shall, in fact, find to be the case.

In order to estimate $\varepsilon_{0}$ we first express it in the form
\begin{equation}
\varepsilon_{0}=\frac{2(seE_{0})^{2}}{V}
\sum_{\vec{k}<k_{u},\lambda}\frac{m[\langle a|\vec{x}|b\rangle
\cdot\hat{\epsilon}_{\lambda}(\vec{k})]^{2}|c_{\vec{k}}|^{2}}{m\omega_{k}^{2}
+e^{2}\rho},
\end{equation}
where
\begin{equation}
c_{\vec{k}}=\sum_{l=1}^{N_{b}}e^{i\vec{k}\cdot\vec{r}_{l}}\sin\theta_{l}\cos
\phi_{l}.
\end{equation}
Taking $k_{u}=2\pi/(\Delta V)^{1/3}$ as the momentum cut off we find that
\begin{equation}
\sum_{\vec{k}<k_{u}}|c_{\vec{k}}|^{2}=N_{b}\sum_{l=1}^{N_{b}}\sin^{2}\theta_{l}
\cos^{2}\phi_{l}\leq N_{b}^{2}.
\end{equation}
This implies that
\begin{equation}
\varepsilon_{0}\leq NE_{0}^{2}|\langle a|\vec{x}|b\rangle |^{2}m.
\end{equation}
Dividing this by $E_{0}N_{b}s$, setting the matrix element equal to the Bohr
radius, and taking $E_{0}$ to be the energy of a 500 nm wavelength photon
we have that
\begin{equation}
\frac{\varepsilon_{0}}{E_{0}N_{b}s}\sim 5\times 10^{-3}.
\end{equation}
Therefore, the minimum should be achieved when $\theta_{l}\cong \pi$.

We can, in fact, do better.  Let us set $\theta_{l}=\pi -\delta\theta_{l}$ and
expand the expression in Eq. (87) in $\delta\theta_{l}$.  We then have that
\begin{equation}
\sum_{l=1}^{N_{b}}E_{0}s\cos\theta_{l}\cong -N_{b}E_{0}s+\frac{E_{0}s}{2}
\sum_{l=1}^{N_{b}}(\delta\theta_{l})^{2},
\end{equation}
and
\begin{eqnarray}
\varepsilon_{0}&\leq &E_{0}^{2}sm|\langle a|\vec{r}|b\rangle |^{2}
\sum_{l=1}^{N_{b}}
(\delta\theta_{l})^{2} \nonumber \\
 & &\leq E_{0}s(5\times 10^{-3})\sum_{l=1}^{N_{b}}(\delta\theta_{l})^{2}.
\end{eqnarray}
From this equation and inequality it is clear that the minimum occurs at 
$\delta\theta_{l}=0$.  Any deviation of $\delta\theta_{l}$ from zero causes
a larger increase in the first term of Eq. (87) than can be compensated for by
the decrease in the second.  
This also implies that the shifts $z_{\vec{k}\lambda}$
are zero and that the operators $a_{\vec{k}\lambda}$ and $b_{\vec{k}\lambda}$ 
are the same.  The lowest order ground state consists of all the spins pointing
down and the field in the vacuum state.  Givien this information the
diagonalization of the rest of the Hamiltonian can proceed as in Section 4.

\section*{Appendix B}
The four eigenvalues of the matrix $GB$ 
are given by $\lambda_{1}=E_{1}(\vec{k})$,
$\lambda_{2}=E_{2}(\vec{k})$, $\lambda_{3}=-E_{1}(\vec{k})$, and
$\lambda_{4}=-E_{2}(\vec{k})$.  The eigenvector of $GB$ corresponding to the 
eigenvalue $\lambda_{j}$ and satisfying Eq. (47) 
as a normalization condition is
\begin{equation}
\eta_{j}=\left(\begin{array}{c} x_{1j} \\ x_{2j} \\ x_{3j} \\ x_{4j}
\end{array} \right),
\end{equation}
where
\begin{eqnarray}
x_{1j}&=&\frac{|\omega_{k}+\lambda_{j}|}{2(\omega_{k}|\lambda_{j}|)^{1/2}}
\left[\frac{(E_{0}^{2}-\lambda_{j}^{2})^{2}}{(E_{0}^{2}-\lambda_{j}^{2})^{2}
+4\omega_{k}E_{0}g_{\vec{k}}^{2}}\right]^{1/2} \nonumber \\
x_{2j}&=& -\frac{2i\omega_{k}g_{\vec{k}}}{(E_{0}-\lambda_{j})(\omega_{k}+
\lambda_{j})}x_{1j} \nonumber \\
x_{3j}&=& \frac{\omega_{k}-\lambda_{j}}{\omega_{k}+\lambda_{j}} x_{1j}
\nonumber \\
x_{4j}&=& -\frac{E_{0}-\lambda_{j}}{E_{0}+\lambda_{j}}x_{2j}.
\end{eqnarray}
Finally, we can use Eq. (45) to give us that
\begin{equation}
M_{ij}=G_{ii}G_{jj}\langle \eta_{i}|c(j)\rangle = G_{ii}G_{jj}x_{ji}^{\ast},
\end{equation}
which, with Eq. (44), implies that
\begin{equation}
A_{ij}=x_{ij}.
\end{equation}

Let us examine, in particular, the matrix elements 
which enter into the calculation
describing two counterpropagating beams (Section 5).  We assume that the beams 
both have wave vectors of magnitude $k_{0}$ and are on the $\alpha$ branch.
Defining 
\begin{equation}
d=\left[\frac{(E_{0}^{2}-E_{1}^{2})^{2}}
{(E_{0}^{2}-E_{1}^{2})^{2}+4\omega_{k_{0}}E_{0}
g_{\vec{k}_{0}}^{2}}\right]^{1/2},
\end{equation}
we have that
\begin{eqnarray}
A_{11} &=& \frac{|\omega_{k_{0}}+E_{0}|d}{2\sqrt{\omega_{k_{0}}E_{1}}}
\nonumber \\
A_{13}&=& \frac{|\omega_{k_{0}}-E_{1}|d}{2\sqrt{\omega_{k_{0}}E_{1}}}
\nonumber \\
A_{21}&=& -\frac{ig_{\vec{k}_{0}}d}{E_{0}-E_{1}}\sqrt{\frac{\omega_{k_{0}}}
{E_{1}}} \nonumber \\
A_{23}&=& \frac{ig_{\vec{k}_{0}}d}{E_{0}+E_{1}}\sqrt{\frac{\omega_{k_{0}}}
{E_{1}}},
\end{eqnarray}
and $A_{31}=-A_{13}$, $A_{33}=-A_{11}$, $A_{41}=A_{23}$, and $A_{43}=A_{21}$.

\pagebreak
\section*{References}
\begin{enumerate}
\item P.\ D.\ Drummond, R.\ M.\ Shelby,S.\ R.\ Friberg, and Y.\ Yamamoto,
Nature \textbf{365},307 (1993).
\item P.\ D.\ Drummond and S.\ J.\ Carter, Rev.\ Mod.\ Phys.\, to be
published.
\item S.\ J.\ Carter, P.\ D.\ Drummond, M.\ D.\ Reid, and R.\ M.\ Shelby,
Phys.\ Rev.\ Lett.\ \textbf{58}, 1841 (1987).
\item P.\ D.\ Drummond and S.\ J.\ Carter, J.\ Opt.\ Soc.\ Am.\ B \textbf{4}
1565, (1987).
\item Y.\ Lai and H.\ A.\ Haus, Phys.\ Rev.\ A \textbf{40}, 844 (1989).
\item H.\ A.\ Haus and Y.\ Lai, J.\ Opt.\ Soc.\ Am.\ B \textbf{7}, 386 (1990).
\item M.\ Rosenbluh and R.\ M.\ Shelby, Phys.\ Rev.\ Lett.\ \textbf{66},
153 (1991).
\item L.\ F.\ Mollenauer, R.\ H.\ Stolen, and J.\ P.\ Gordon, Phys.\ Rev.\ 
Lett.\ \textbf{45},1095 (1980).
\item M.\ Hillery and L.\ Mlodinow, Phys.\ Rev.\ A \textbf{30}, 1860 (1984).
\item P.\ D.\ Drummond, Phys.\ Rev.\ A \textbf{42},6845 (1990).
\item B.\ Huttner and S.\ M.\ Barnett, Phys.\ Rev.\ A \textbf{46}, 4306 (1992).
\item I.\ Deutsch, Ph.\ D.\ thesis University of California at Berkeley, 
unpublished (1992).
\item M.\ Hillery and L.\ Mlodinow, Phys.\ Rev.\ A \textbf{31}, 797 (1985).
\item L.\ Mlodinow and N.\ Papanicolaou, Ann.\ Phys.\ (NY) \textbf{128},
314 (1980).  For a review see A. Chaterjee, Phys.\ Rep.\ \textbf{186}, 
250 (1990).
\item J.\ J.\ Hopfield, Phys.\ Rev.\ \textbf{112}, 1555 (1958).
\item T.\ Holstein and H.\ Primakoff, Phys.\ Rev.\ \textbf{58}, 1098 (1940).
\end{enumerate}
\end{document}